# Strain heterogeneity and micro-damage nucleation under tensile stresses in an Mg-5Al-3Ca alloy with an intermetallic skeleton

M. Zubair[a,b], S. Sandlöbes-Haut[a], M.A. Wollenweber[a], K. Bugelnig[c], C.F. Kusche[a], G. Requena[c], S. Korte-Kerzel[a]

[a] Institute for Physical Metallurgy and Materials Physics, RWTH Aachen University, Kopernikusstr. 14, 52074, Aachen, Germany
[b] Department of Metallurgical and Materials Engineering, G.T Road, UET Lahore, Pakistan
[c] German Aerospace Centre, Linder Höhe, 51147 Cologne, Germany

**Abstract**

Strain heterogeneity at the microstructural level plays a vital role in the deformation and fracture behaviour of dual or multi-phase materials. In the present work, the strain heterogeneity, localization and partitioning arising at the sub-micron scale during elevated temperature (170 °C) tensile deformation of an Mg-5Al-3Ca alloy was investigated using quasi in-situ µ-DIC experiments. The results reveal that the strain is mainly carried by the α-Mg phase, while the intermetallic Laves phase plays a critical role in that strain concentrations build up at the α-Mg matrix and Laves phase interfaces, hence, reducing the overall deformability of the alloy. In quasi in-situ and bulk material analysis at elevated temperature, cracks were observed to nucleate in the Laves phase, at i) the intersection points of slip lines in the α-Mg matrix with the Laves phase and ii) the twin intersections with α-Mg/Laves phase interfaces and iii) twin transmissions across α-Mg/Laves phase interfaces. Euler number analysis has shown that the (inter-)connectivity of the Laves phase decreases with deformation. Finally, cracks grow preferentially along the Laves phases until the material fractures.

1. Introduction

During the last years, Mg-based alloys have attracted attention of many researchers around the globe particularly due to their low density, low cost and high availability [1, 2]. Al is the main alloying element of most commercial Mg alloys (Al-Zn or Al-Mn) as it enhances the strength and hardness and also the corrosion resistance and castability [2, 3]. However, the high temperature mechanical properties of Mg-Al alloys are poor, mainly because of the presence of $Mg_{17}Al_{12}$ as intermetallic precipitate phase in the microstructure. $Mg_{17}Al_{12}$ has a low thermal stability and readily softens at temperatures above 130 °C [4-6].

Ca addition to Mg-Al alloys can considerably enhance the elevated temperature mechanical properties like yield strength [7], tensile strength, creep properties [8-11] and thermal stability [4]. This is because Ca addition suppresses the formation of the $Mg_{17}Al_{12}$ phase with a low thermal stability and promotes the formation of harder Laves phases with a higher thermal stability. At low concentrations, Ca tends to dissolve in the matrix and in the second phase ($Mg_{17}Al_{12}$), thereby, enhancing its thermal stability [4]. Once the Ca concentration exceeds a critical value, the Laves phase starts to appear in the microstructure. Specifically, Kondori et al. [4] observed i) only $Mg_{17}Al_{12}$ as second phase in AM60 and modified AM60 containing 0.5 wt.% Ca, ii) a combination of $Mg_{17}Al_{12}$ and $Al_2Ca$ Laves phase in modified AM60 with 1.2 wt.% Ca addition, and iii) only $Al_2Ca$ Laves phase when the Ca content was 2.0 wt.%. The two other Laves phases of the Mg-Al-Ca system, namely, $(Mg,Al)_2Ca$ and $Mg_2Ca$, were also reported to form in Mg-Al-Ca alloys [12-15]. It was reported that the Ca/Al ratio controls which Laves phases form: the type of Laves phase changes from $Al_2Ca$ (C15) to $(Mg,Al)_2Ca$ (C36) and $Mg_2Ca$ (C14) with variation of the Ca/Al ratio from 0.3 to 1.0 [12, 16].

The alloy Mg-5Al-3Ca (in wt.%) (from here onwards referred to as AX53) has a dual phase microstructure, containing a hard intermetallic skeleton of $(Mg,Al)_2Ca$ Laves phase embedded in a soft α-Mg matrix [16]. Due to significantly different intrinsic mechanical properties of the different phases (α-Mg matrix and Laves phases) [17-19], a highly heterogeneous strain distribution was reported for Mg-Al-Ca alloys [16, 20]. In addition, the volume fraction, morphology, distribution, and orientation relationship between matrix and second phase can affect the strain heterogeneity. Strain heterogeneity usually results in strain partitioning, localization and shear band formation, thus significantly affecting the mechanical and fracture behaviour of dual or multi-phase materials [21-23]. Analysis of the strain distribution and partitioning from the grain level down to the sub-micrometer scale can provide valuable insights regarding the mechanical and fracture behaviour of multi-phase Mg-Al-Ca alloys.

Several recent studies focused on the evaluation of grain level strain heterogeneities in multi-phase materials [21, 23, 24]. Digital Image Correlation (DIC) is an advanced method to quantitatively measure the strain partitioning occurring at macroscopic and microscopic scales under complex stress states [22, 24-35]. In DIC, images of the same sample area before and after straining are correlated by measuring and quantifying the local relative displacements before and after deformation. In order to quantify the relative displacements, speckles or small particles are deposited on the sample surface and the relative displacement of the speckles are correlated for the deformed and un-deformed image [28]. The local relative displacements are commonly considered to result from strain in the microstructure. A small speckle size or use of nanoparticles such as $SiO_2$ or $Al_2O_3$ enables to measure the local strain at a sub-micrometer scale using scanning in (SEM) as demonstrated e.g. by Dutta et al. [21] for a medium Mn steel and Zubair et al. [16] for Mg-Al-Ca alloys.

In the present work, the strain distribution and partitioning between the α-Mg matrix and the (Mg,Al)$_2$Ca Laves phase evolving during deformation in the dual phase AX53 alloy was investigated using high resolution sub-micrometer scale quasi in-situ experiments coupled with DIC and electron backscatter diffraction (EBSD). The tensile and fracture behaviour of alloy AX53 was also studied using quasi in-situ and ex-situ deformation experiments in SEM. To evaluate any changes in the connectivity of the Laves phase network during deformation, the Euler number was used as a mathematical method to express the connectivity of objects and to measure the effect of deformation on the (loss of) connectivity [36].

## 2. Experimental Methods

Pure Mg, Al and Ca were molten and solidified in an induction furnace with a pressure of 15 bar Argon using a steel crucible. The global chemical composition of the as-cast alloy was measured using wet chemical analysis, Table 1.

*Table 1: Composition of the as cast alloy*

| Alloy | Al (wt. %) | Ca (wt. %) | Fe (wt.%) | Mg (wt.%) |
|---|---|---|---|---|
| AX53 | 5.21 | 3.18 | 0.008 | balance |

Samples for microstructure analysis were metallographically prepared by mechanical grinding and polishing, electrolytic polishing and a subsequent final mechanical polishing step. First, the samples were mechanically ground down to 4000 SiC emery paper and then mechanically polished using 3 µm and 1 µm diamond suspension. Subsequently, the samples were electro-polished using the AC2 (Struers) electrolyte (electro-polishing was needed to prepare good surfaces for EBSD analysis). The temperature, time and voltage of electro-polishing were ≤ -20 °C, 60 s and 15 V respectively. Due to the different electrochemical behaviour of Mg and the Laves phase, electro-polishing produced a waviness at the sample surface. Finally, this waviness was removed via mechanical polishing using OPU (~ 40 nm $SiO_2$ colloidal suspension) followed by cleaning in an ultrasonic bath and softly rotating on a clean cloth using ethanol as a cleaning agent. For µ-DIC, the sample was not cleaned and the residual nano-sized $SiO_2$ particles acted as speckles (a necessary prerequisite for strain partitioning analyses using DIC). An acceleration voltage of 10-20 kV was used to characterize the as-cast and deformed microstructures as described in detail in [16]. Energy dispersive X-ray spectroscopy (EDS) was carried out at the lower voltage of 10kV and the higher voltage of 20 kV was used for secondary electron (SE) imaging, back-scattered electron (BSE) imaging and electron backscatter diffraction (EBSD). The images and EBSD data were analysed using ImageJ, Channel 5 and OIM. In order to investigate large microstructural areas at a sub-micrometer scale resolution, panoramic imaging was extensively used within this study for µ-DIC and fracture analysis. The software used for image stitching was Image Composite Editor (ICE) [37].

For the evaluation of the change in local connectivity of the 2D Laves phase network as a function of composition and deformation, SEM images of the same sample positions before and after tensile deformation were used. Images taken on the sample prepared in this work were further compared with samples possessing different Ca/Al ratios produced in the same way and discussed in terms of their mechanical properties in our previous work [16]. Images were smoothed with a bilateral filter and the 2D networks were segmented using global gray value thresholds. The segmented networks were then labeled so that a different color was assigned to each individual Laves phase segment. Afterwards, the local connectivity of the 2D segmented Laves networks, i.e. the amount of connecting branches of the network, was quantitatively evaluated using the software Avizo Fire 9.5 by means of the Euler number, χ as a topological parameter. [36, 38]. This parameter reveals changes in local connectivity in complex network structures. In two dimensions, the Euler number can be calculated as follows:

$$\chi = b_0 - b_1 \qquad (1)$$

where $b_0$ is the number of objects and $b_1$ the number of holes within the object [39]. Variations in the Euler number indicate a change in local connectivity. More specifically, a decreasing Euler number indicates an increase of local connectivity and vice versa [36, 38].

An electrochemical testing machine (ETM) was used to carry out tensile deformation of standard dog bone shaped specimens having a gauge length of 10 mm at room temperature (RT) and 170 °C. An initial strain rate of $5\times10^{-4}$ s$^{-1}$ was selected for tensile testing at room temperature and 170 °C. A linear variable differential transformer (LVDT) extensometer was used to measure the elongation of the specimens during tensile deformation. Quasi in-situ µ-DIC experiments were done using SEM (Zeiss LEO1530) to study the distribution of strain at the sub-micrometer level during tensile deformation at 170 °C. The sample was deformed stepwise at 170 °C in an ETM and monitored using SEM after certain pre-defined deformation steps of 3, 4 and 5% global strain. The local strain was determined using the ARAMIS GOM Correlate software. For deformation and fracture analysis, the samples were deformed both, quasi in-situ and ex-situ.

3. Results

3.1 Microstructure Analysis

The microstructure of as-cast AX53 is shown in Fig. 1 (a). EDS point analysis of the Laves phases was performed as highlighted by the characters A – D in Fig. 1 (b) and the corresponding compositions are listed in Table 2.

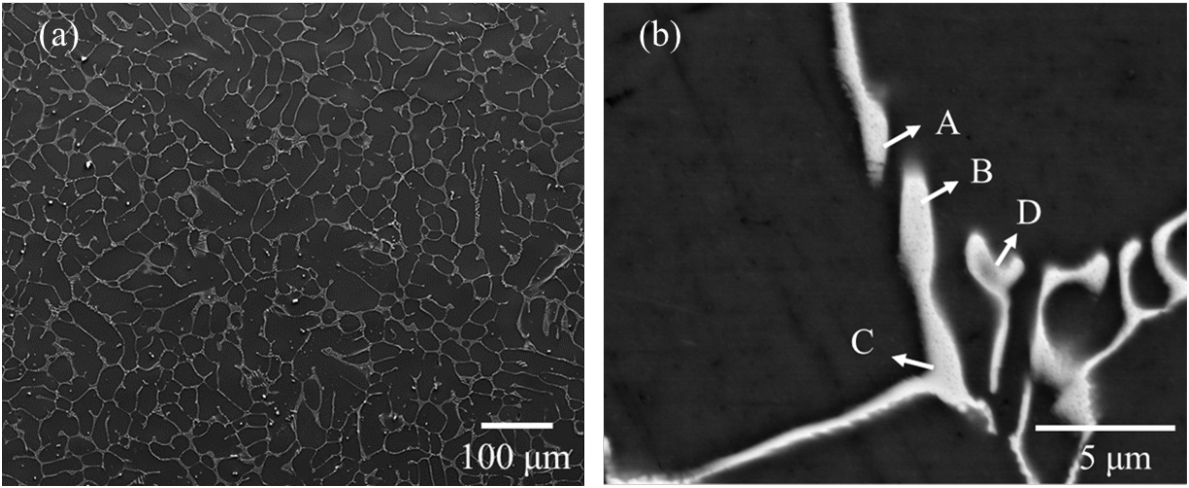

*Fig. 1 (a): SE micrograph of the microstructure of as-cast alloy AX53, (b) high resolution BSE image of the alloy. Compositional analysis of the Laves phases via EDS was performed on the points marked by the characters A, B, C and D, see Table 2.*

*Table 2: Compositional analysis of the Laves phases in alloy AX53 using EDS.*

| Spots | Mg (at. %) | Al (at. %) | Ca (at. %) | Corresponding Phase |
|---|---|---|---|---|
| **A** | 55.3 | 27.7 | 17.1 | $(Mg,Al)_2Ca$ |
| **B** | 33.6 | 40.3 | 26.2 | $(Mg,Al)_2Ca$ |
| **C** | 32.7 | 40.6 | 26.7 | $(Mg,Al)_2Ca$ |
| **D** | 64.9 | 22.6 | 12.5 | $(Mg,Al)_2Ca$ |
| **Matrix (α-Mg)** | 98.3 | 1.7 | - | - |

The area fraction of intermetallic phase amounts to around 7.3 % and the mean dendrite cell size (as calculated using the linear intercept method, ASTM E 112–12 ) is 28.5 ± 3.4 μm in the as cast alloy. The grain size of the α-Mg matrix is much larger than the dendrite cell size as can be seen from the inverse pole figure (IPF) map in Fig. 2. The average grain size of the α-Mg matrix is 183.5 ± 72.5 μm (determined using the same procedure i.e. according to ASTM E 112-12).

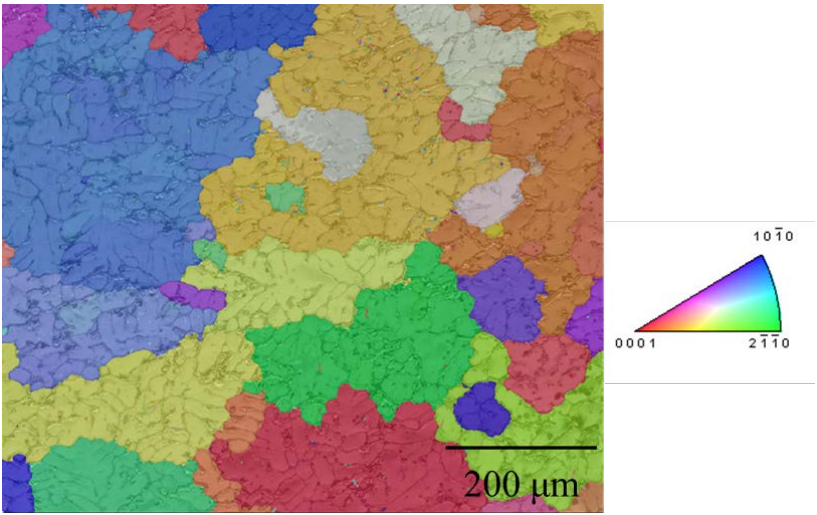

*Fig. 2: Superimposed IPF (inverse pole figure) and band contrast (BC) maps of the as-cast alloy showing that the dendrite cell size is much smaller than the α-Mg grain size.*

## 3.2 Tensile Properties

Fig. 3 a) represents the true stress-strain curve of the as-cast alloy at RT and 170 °C. With increasing temperature from RT to 170 °C, the yield strength ($\sigma_{0.2}$) drops from 92 MPa to 74 MPa, the uniform elongation (UE) increases from 1.2 to 6.2 %, and the ultimate tensile strength (UTS) stays nearly constant (≈ 123 MPa at RT and ≈ 122 at 170 °C).

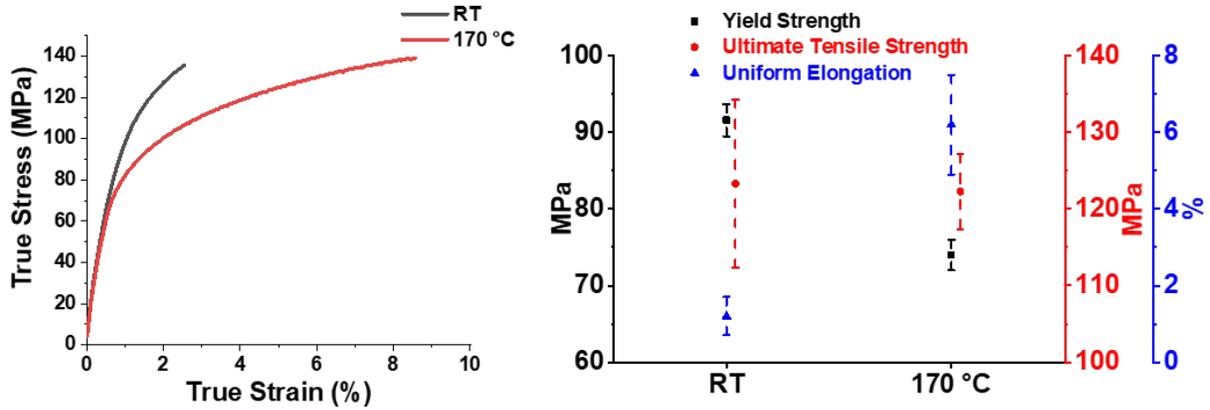

*Fig. 3 (a) True stress-strain curve of the as-cast alloy deformed at RT and 170 °C, (b) Comparison of yield strength ($\sigma_{0.2}$), ultimate tensile strength and uniform elongation at RT and 170 °C.*

### 3.3 Strain partitioning

As shown in Fig. 1, the microstructure of the as-cast alloy contains two phases: a soft α-Mg phase reinforced with a hard interconnected Laves phase (predominant network of $(Mg,Al)_2Ca$). In order to understand the deformation behaviour of alloy AX53, it is of prime interest to investigate how strain is distributed at the sub-micrometer level. For that purpose, quasi in situ µ-DIC in conjunction with EBSD mapping was performed. In order to achieve a high resolution DIC map of a statistically relevant large sample area, panoramic imaging was used. For panoramic imaging, 56 individual images of 33.0 µm by 24.8 µm and a magnification of 3.46 KX (7(rows) x 8(columns), with 20 % overlap each) were captured and stitched together.

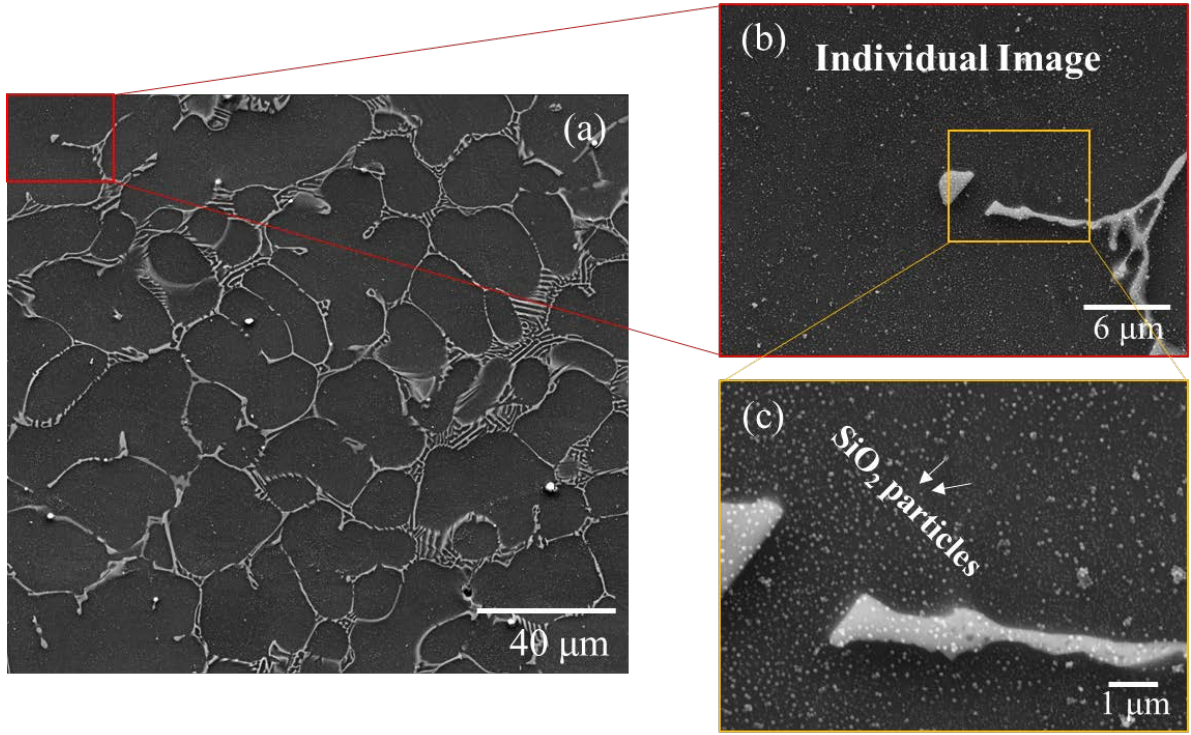

*Fig. 4. (a) Un-deformed panoramic image, (b) one individual image of magnification 3.46 KX and (c) High magnification image showing the speckle patterns on the sample surface.*

Fig. 4 (a) depicts the stitched panoramic image of the undeformed specimen. One of the individual images of which the panorama is composed of, is shown in Fig. 4 (b) and (c) displays the $SiO_2$ nano-particle speckles on the sample surface.

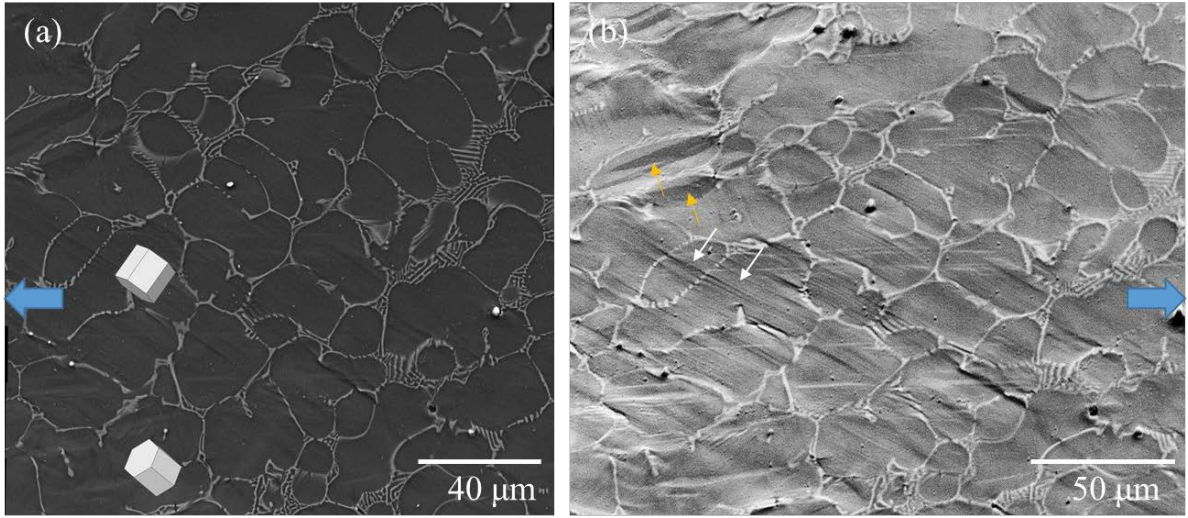

*Fig. 5. (a) SE panoramic image of the same sample surface region (as shown in Fig. 4 (a)) after quasi in-situ straining to a global strain of 4% at 170 °C and (b) deformed image at a stage tilt of 70 °. The blue arrows indicate the tensile direction, white arrows show slip lines and yellow arrows highlight twins. The unit cells in (d) show the orientation of the α-Mg matrix (determined from EBSD of deformed microstructural regions).*

The deformed microstructure of the same sample area after 4 % global straining at 170 °C is shown in (d) and (e). The unit cells in Fig. 5 (a) indicate the orientation of the α-Mg matrix. Basal slip lines (highlighted by white arrows in Fig. 5 (b)) and tensile deformation twins (highlighted by yellow arrows in Fig. 5 (b)) are observed in the deformed microstructure.

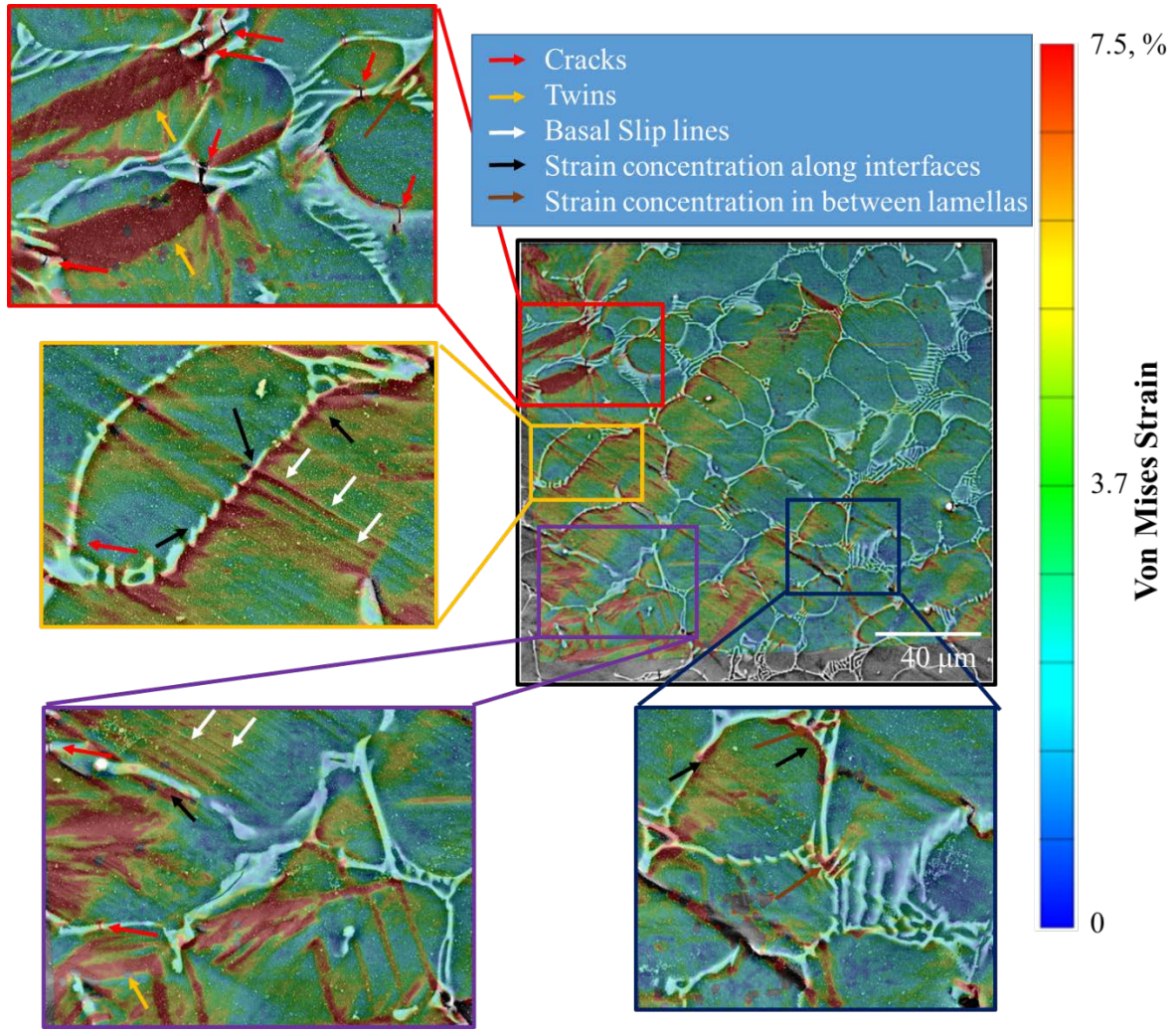

*Fig. 6. Von Mises strain map (facet size: 0.55 µm, step size: 0.44 µm) of the deformed microstructure region depicted in Fig. 4. Laves phase cracks are marked in the insets with red arrows, twinned regions with yellow arrows, basal slip lines with white arrows, strain concentrations along α-Mg/Laves phase interfaces with black arrows and strain concentration in between the eutectic Laves lamellas with orange arrows.*

Fig. 6 shows a 2D von Mises strain map of the microstructural region given in Fig. 4 and Fig. 5. The map reveals highly heterogeneous strain at the sub-micron level in the microstructure. The strain tends to concentrate (i) along basal slip traces (white arrows in Fig. 6), (ii) at deformation twins (yellow arrows), (iii) along α-Mg/Laves phase interfaces (black arrows), and (iv) in between the eutectic Laves phase lamellas (orange arrows). In the Laves phase, cracks are visible in the regions of high strain concentrations and intersections of slip traces and deformation twins with α-Mg/Laves phase interfaces (red arrows in Fig. 6). Subsequently, the sample was also deformed to 5.5% global strain, however, the strain map was similar to the one after 4% global strain with an increased surface topography limiting detailed microstructure observations and is therefore not shown.

The Euler number was used to describe the evolution of the connectivity of the Laves phase skeleton during deformation. For this purpose, the Laves phase network before and after deformation was first segmented, where each segment represents an interconnected Laves phase structure, Fig. 7. The different colours correspond to different Laves phase segments.

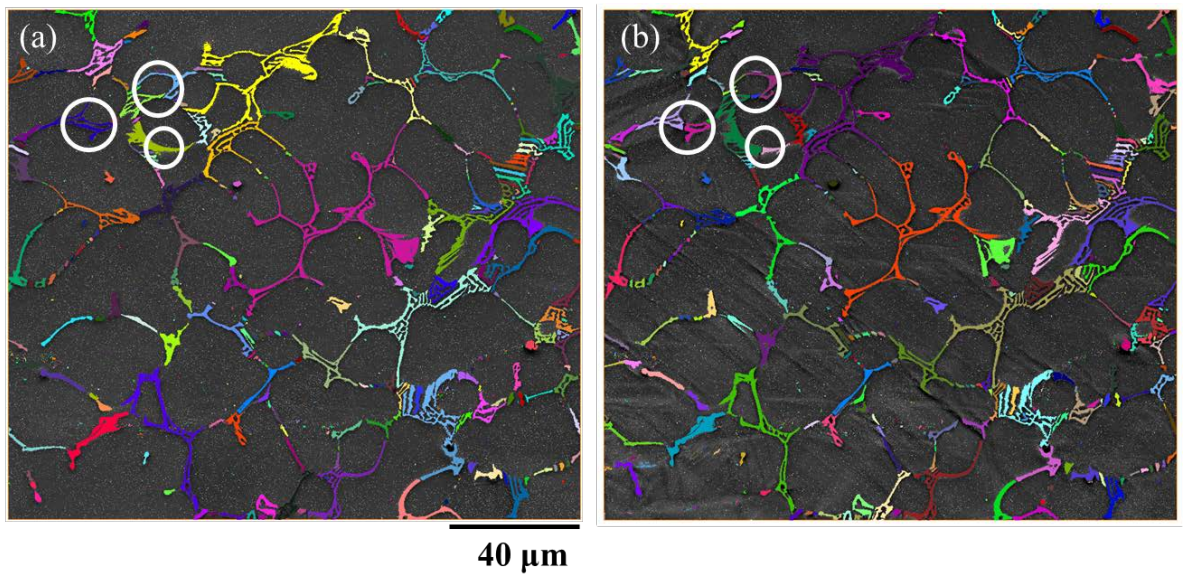

*Fig. 7. Segmented Laves phase network (a) before deformation, Euler number: -1363, (b) after 4% global strain at 170 °C, Euler number: -594. White circles indicate fragmentation of the Laves phase network during deformation.*

From Fig. 7 it is evident that deformation leads to fragmentation of the Laves phase network visible as an increased number of Laves phase segments in the deformed microstructure and a correspondingly quantified by an increased Euler number.

### 3.4 Deformation and Fracture Behavior

Fig. 8 (a) reveals the microstructure of the alloy after quasi in-situ deformation at 170 °C to 3% global strain. Fig. 8 (b) and (c) depict magnified micrographs of a region undergoing basal slip and tensile twinning. Basal slip lines are highlighted by white arrows in the Fig. 8 (b and c). In addition to basal slip, traces of $(10\bar{1}1)$ pyramidal slip are observed and highlighted by blue arrows. Further, it can be seen that at the intersection points of basal slip traces in the α-Mg matrix with α-Mg/Laves phase interfaces, small cracks are formed. Some cracks are also visible at intersections of deformation twins with α-Mg/Laves phase interfaces. Fig. 8 (d), (e) and (f) show the IPF and Schmid factor maps for basal slip and tensile twinning superimposed on BC maps. Basal slip and tensile twinning is indeed observed in grains with high Schmid factors for the corresponding deformation mechanisms, viz. basal slip and tensile twinning.

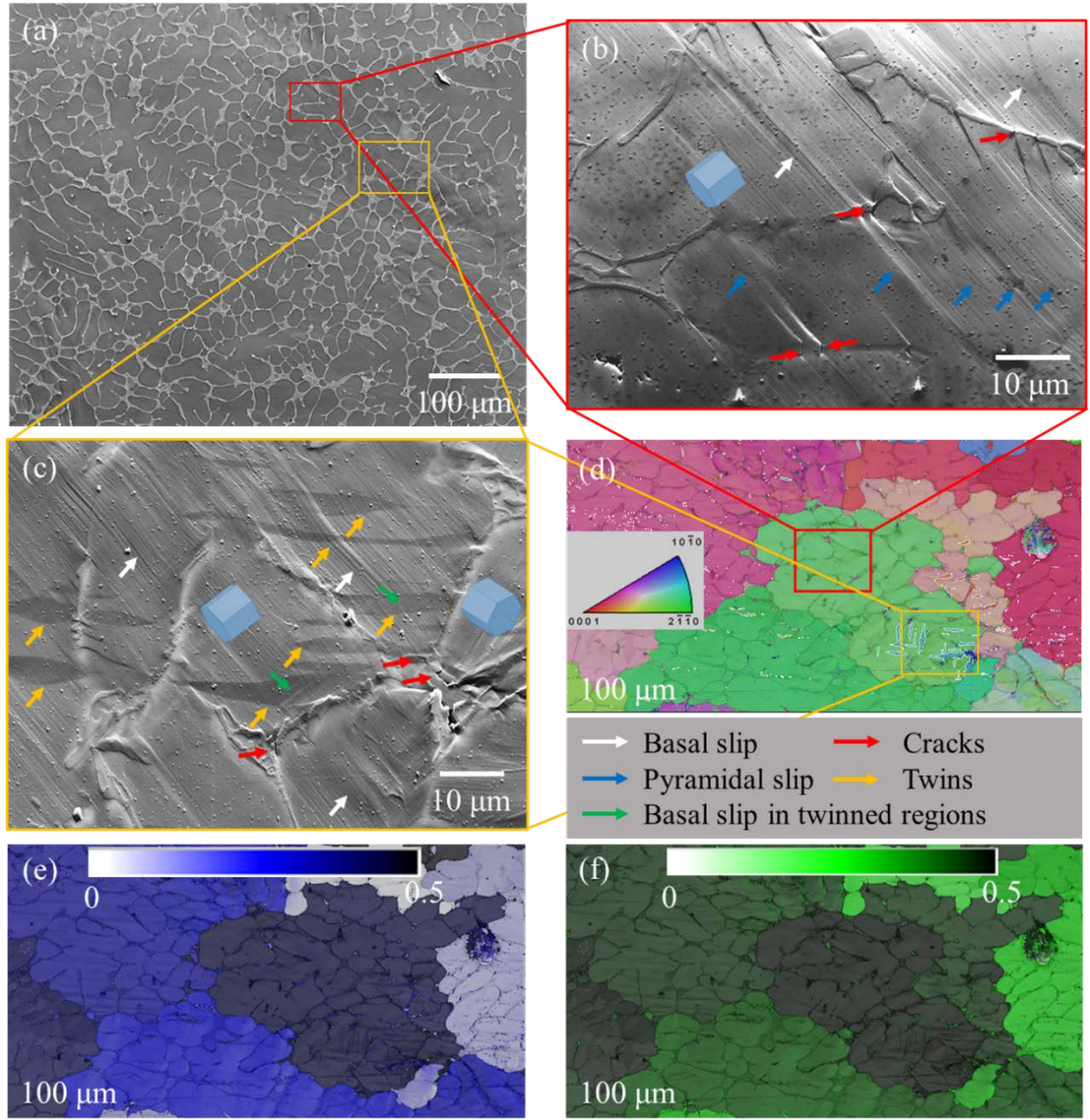

*Fig. 8. (a) SE image of a sample after 3% global strain (quasi in-situ deformation) at 170 °C, (b) and (c) SE images of the regions highlighted by yellow and red boxes in (a) at a stage tilt of 70 °. The α-Mg matrix orientation is depicted by the unit cells, white arrows highlight basal slip traces, red arrows cracks in the Laves phase, blue arrows $(10\bar{1}1)$ pyramidal slip, yellow arrows tensile deformation twins and green arrows basal slip lines in the twinned regions. (d) Superimposed IPF and BC maps, tensile twins are highlighted by white lines. (e) Schmid factor map for basal slip and (f) Schmid factor map for tensile twinning; the area shown in (e) and (f) is identical to the IPF map shown in (d).*

To verify that the formation of cracks observed at the sample surface during quasi in-situ deformation (Fig. 6 and Fig. 8) is not an artifact induced by the free surface of the sample, characterisation of the bulk of deformed samples was performed in addition to the quasi in-situ tests. As the sample was metallographically prepared after deformation, no slip traces are visible, but only deformation twins. Specifically, Fig. 9 (a) shows superimposed BC and IPF maps of the microstructure after deformation to 4% global strain at 170 °C and (b) shows cracks nucleating in the Laves phase at twin – α-Mg/Laves phase interfaces intersection points.

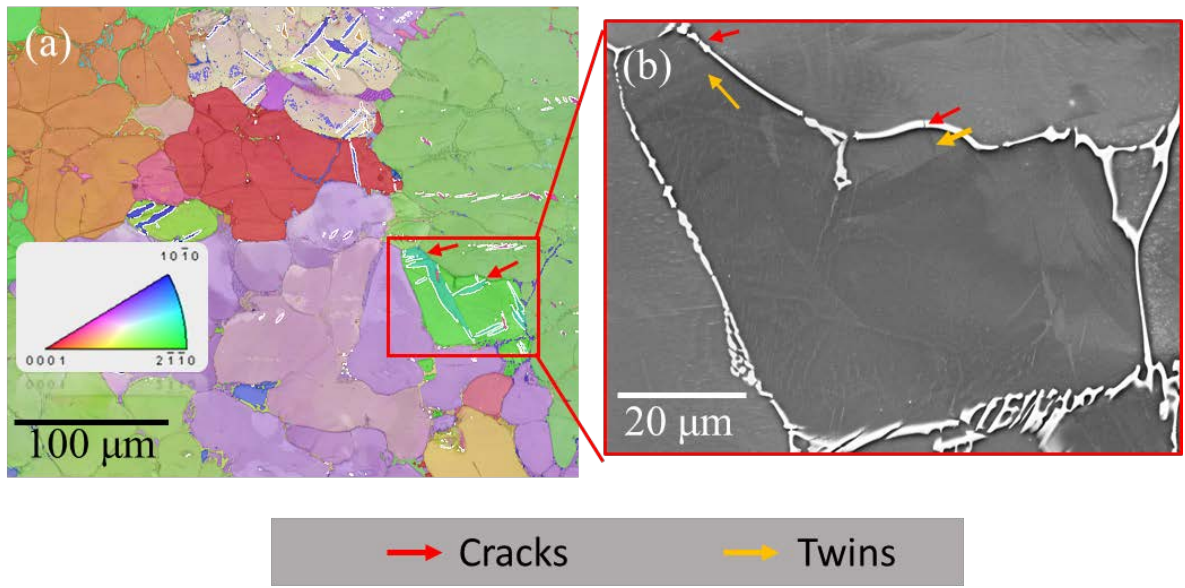

*Fig. 9. (a) Superimposed IPF and BC maps from the bulk of a sample after 4% deformation at 170 °C, (b) SE image highlighting cracks at twin intersection points (red arrows) in the Laves phases of the sample shown in (a).*

Micrographs of samples quasi in-situ deformed at 170 °C to 5% global strain are given in Fig. 10 (a) and (b). Cracks were observed to form at the intersection points of basal slip traces in the Mg matrix with α-Mg/Laves phase interfaces (Fig. 10 (c)) and at points where deformation twins transfer from the Mg matrix through the Laves phase (Fig. 10 (b)).

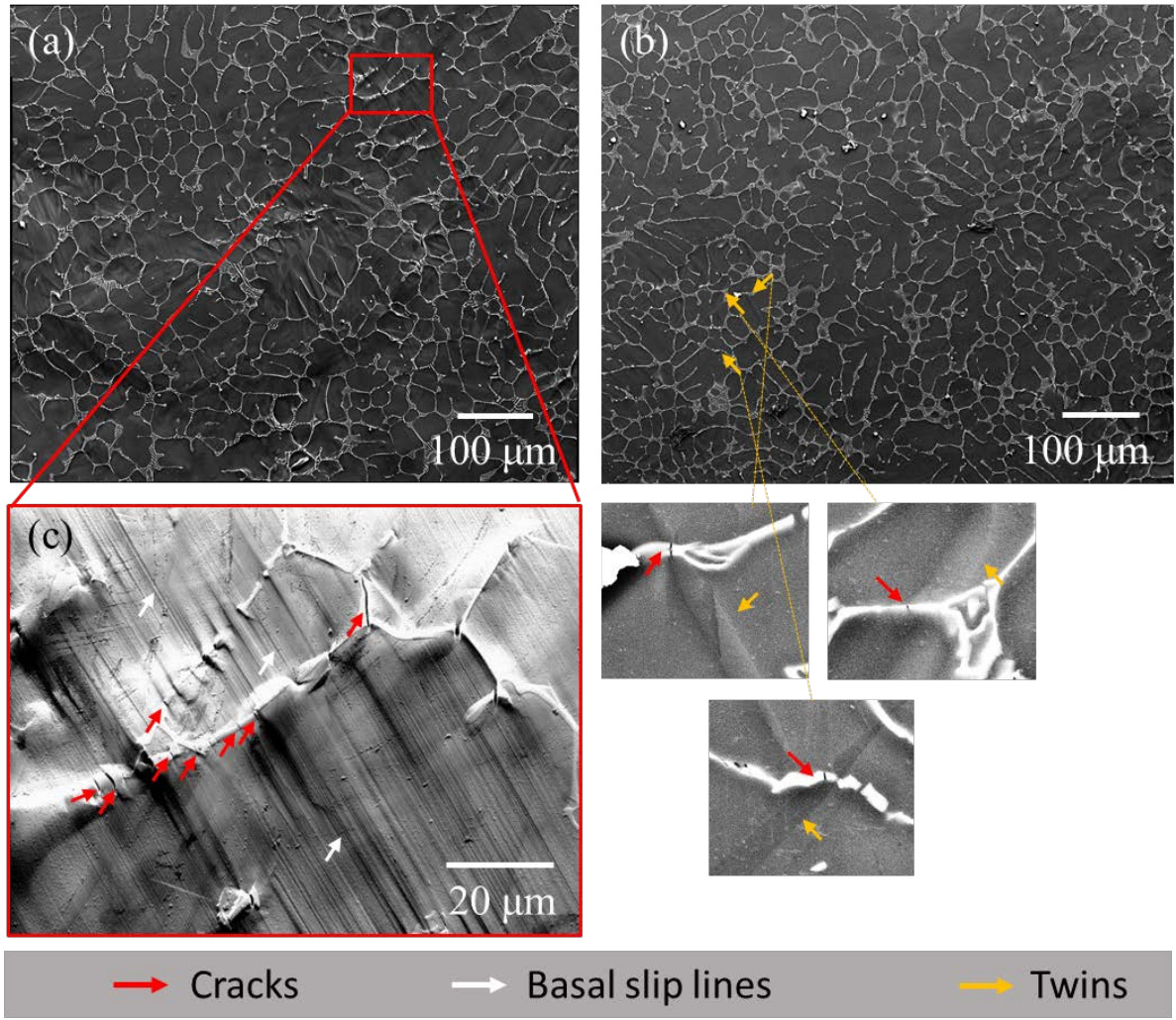

*Fig. 10. (a) and (b) SE images of the sample at 5% global strain (quasi in-situ) at 170 °C, (c) depicts the SE image of the region in inset in (c) with stage tilted at 70 ° to the electron beam. Cracks nucleation points in the Laves phase are shown by red arrows, basal slip lines as white arrows, yellow arrows in the magnified portion of (b) represent twins and red arrows depict cracks in the Laves phase.*

## 4. Discussion

### 4.1 Microstructure and mechanical properties

The microstructure of as-cast AX53 consists of the two phases α-Mg and (Mg,Al)$_2$Ca Laves phase. Luo et al. [40] have shown that these commonly share an orientation relationship of (001)Mg//(001)(Mg,Al)$_2$Ca and [010]Mg//[110](Mg,Al)$_2$Ca. Similar microstructures were also observed before in other Mg-Al-Ca alloys [16, 20, 40]. The average dendrite cell is slightly lower and the area fraction of Laves phase slightly higher than the values reported for a similar Mg-Al-Ca alloy (with a Ca/Al ratio of 0.6) in our previous work [16]. On the other hand, the $\sigma_{0.2}$ and UTS of the present alloy at both temperatures (RT and 170 °C) are significantly higher. Specifically, the $\sigma_{0.2}$ at RT of the present alloy is 30 % higher and the $\sigma_{0.2}$ at 170 °C is 26.5% higher, while the corresponding UTS values are 13.5% higher at RT and 21.7% higher at 170 °C. This is assumed to be due to different melting and casting conditions: the alloy investigated in the current study was molten and cast under 15 bar Ar pressure, whereas the one studied in [16] was molten and cast under ambient conditions. Melting and casting under 15 bar Ar pressure is assumed to cause a lower number of casting defects than at ambient conditions resulting in better mechanical properties.

### 4.2 Strain localisation and micro-damage

As shown in Fig. 6, it is generally observed that the α-Mg grains carry most of the imposed strain, while the Laves phase does not undergo a significant amount of plastic deformation. Specifically, von Mises strains of about 0.5-1.5% were observed, considerably lower than the applied global strain of 4%, Fig. 6. Further, strain localization was observed i) along basal slip traces, ii) at deformation twins, iii) at α-Mg/Laves phase interfaces, and iv) in between the eutectic Laves phase lamellas. These can be categorized into two distinct types of strain localisation:

**Type 1: Strain partitioning and localisation in the α-Mg matrix**

The α-Mg matrix deforms mainly by basal slip and tensile twinning, Fig. 5, 8-10. It can be further seen in Fig. 6, that some grains carry high strain while other grains exhibit only relatively low strain levels. Strain localisation in those grains which carry high strain levels occurs preferentially along basal slip traces and tensile twins. This is assumed to be due to the different grain orientations and the plastic anisotropy of Mg. Specifically, in pure Mg, the critical resolved shear stresses (CRSS) for basal slip (~0.52 MPa [41]) and tensile twinning (2.4 MPa [42]) are low when compared to prismatic slip ($\approx$ 20-39 MPa [43-45]) and pyramidal slip ($\approx$ 44 MPa [44, 46]). Although the exact CRSS values depend on temperature and composition, the trend of low CRSS for basal slip and tensile twinning and high CRSS for prismatic and pyramidal slip was reported for a huge number of Mg alloys in the temperature range RT-170 °C [45, 47-49]. Consequently, only grains with a high Schmid factor for either basal <a> slip or tensile twinning, undergo significant amounts of deformation through slip and twinning (see Fig. 8). Therefore, strain localisation in α-Mg is proposed to be primarily dependent on the crystallographic orientations of the grains.

**Type 2: Strain localisation along the α-Mg/Laves phase interface**

In addition to strain concentrations in the α-Mg grains, strain localization was observed along the α-Mg/Laves phase interfaces and in between the eutectic Laves phase lamellas, Fig. 6. As no preference regarding the crystallographic orientation and misorientation is observed, it is assumed that these strain localisations are caused by grain/phase boundary sliding during deformation at 170 °C. The occurrence of grain and phase boundary sliding during elevated temperature deformation and creep in Mg-Al-Ca alloys was also reported in our own previous work and by several other researchers [7, 16, 40, 50]. This type of strain partitioning and localization does not seem to have any geometrical or orientation dependence. However, higher strain accumulation are observed at interfaces adjacent to α-Mg grains that carry high strain, which is proposed to be caused by the higher amount of strain induced on the interface through larger deformation of the respective α-Mg grains.

**Cracks and micro damage**

In line with our previous work and the reports of other groups [10, 16, 51], crack nucleation was observed in the Laves phase in the present study. More detailed analysis based on quasi in-situ deformation experiments in the present study showed that these crack nucleation sites are not random but correspond to strain concentration sites, Fig. 6, mainly at the intersection points of basal slip traces and deformation twins with α-Mg/Laves phase interfaces. Similarly, Bieler et al. [52] showed that, in single phase materials, damage tends to nucleate at grain boundaries that are unfavorable for slip transmission. In many alloys, hard second phases or interfaces (between hard and soft phase) were reported to act as preferential damage nucleation sites [10, 50, 51, 53-55]. The observation of preferred crack nucleation at slip band intersections with α-Mg/Laves phase interfaces, Fig. 8 (b) and Fig. 10 (c), indicate that stress concentrations are generated at the intersections due to the pile-up of dislocations. Dislocation pile-ups at the α-Mg/Laves phase interfaces are assumed to be prevalent as the hard Laves phase skeleton is considered as an effective obstacle to dislocation movement [9, 19]. A more detailed large-scale analysis of the crack nucleation sites revealed that those microstructural regions where the α-Mg matrix carries a considerable amount of strain induced by dislocation slip and /or twinning contain also a high amount of cracks in the Laves phase. In contrast, those α-Mg grains that do not carry much strain, exhibit far fewer cracks. In Fig. 11 (a), cracks that developed during successive deformation are highlighted by yellow (slip intersections points), blue (twin intersection points) and orange circles (where no distinction between slip or twin intersection can be made). Fig. 11 (b), (c) and (d) represent the IPF map and the Schmid factor maps for basal slip and tensile twinning superimposed on BC maps. It is evident from these micrographs that the α-Mg orientation determines the amount of cracks that formed during deformation. Specifically, grains with a high Schmid factor for basal slip or tensile twinning contain also a high number of cracks predominantly at intersections of slip lines and twins with α-Mg/Laves phase interfaces. Fig. 11 (e) also indicates that the number of cracks appearing at slip intersection points are much more than twin intersection points with α-Mg/Laves phase interfaces. This is primarily because the number of slip lines intersections is significantly higher than twin intersections as basal slip is the predominant deformation mechanism observed.

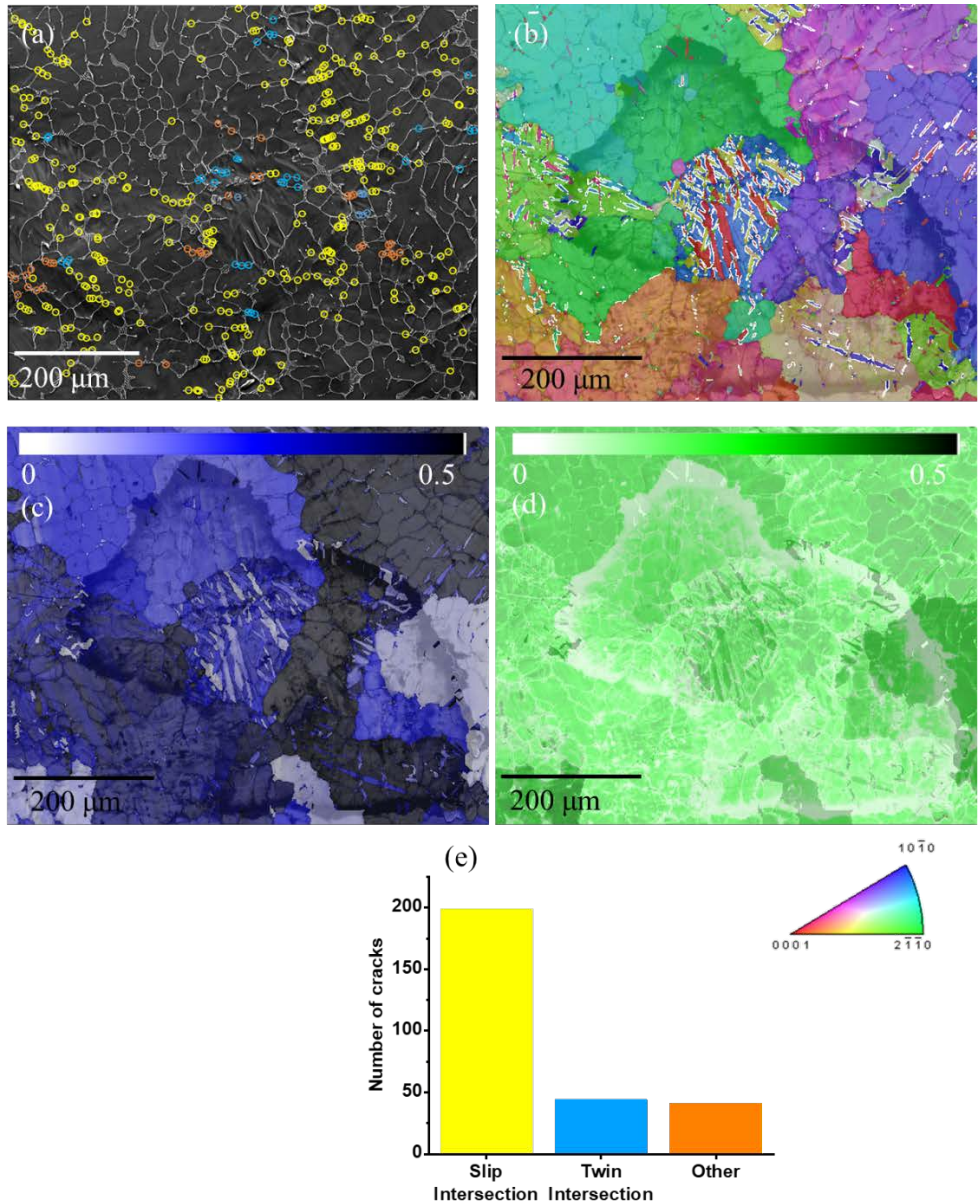

*Fig. 11. (a) SE image of deformed microstructure after 5% global deformation at 170 °C, cracks at slip intersection points are highlighted by yellow circles, those at twin intersections by blue circles and others by orange circles, (b) Superimposed IPF and BC maps of approximately the same region shown in (a), tensile twin boundaries are highlighted by white lines. (c) Schmid factor map for basal slip and (d) Schmid factor map for tensile twinning; the area shown in (c) and (d) is identical to the IPF map shown in (b). (e) depicts the fraction of cracks appearing at slip band intersection (SI) and twin intersection (TI) with α-Mg/Laves phase interfaces, "others" are cracks which could not be related to a slip band – interface or twin – interface intersection.*

Similarly, crack nucleation at intersections of twins with grain boundaries and at twin boundaries was reported before for Mg and Ti alloys [56-58]. With continued straining, some of these cracks eventually grow preferentially along the Laves phase network and cause fracture of the material. Fig. 12 shows a micrograph of a sample deformed at 170 °C until fracture in which the crack causing failure propagated preferentially along the Laves phase network.

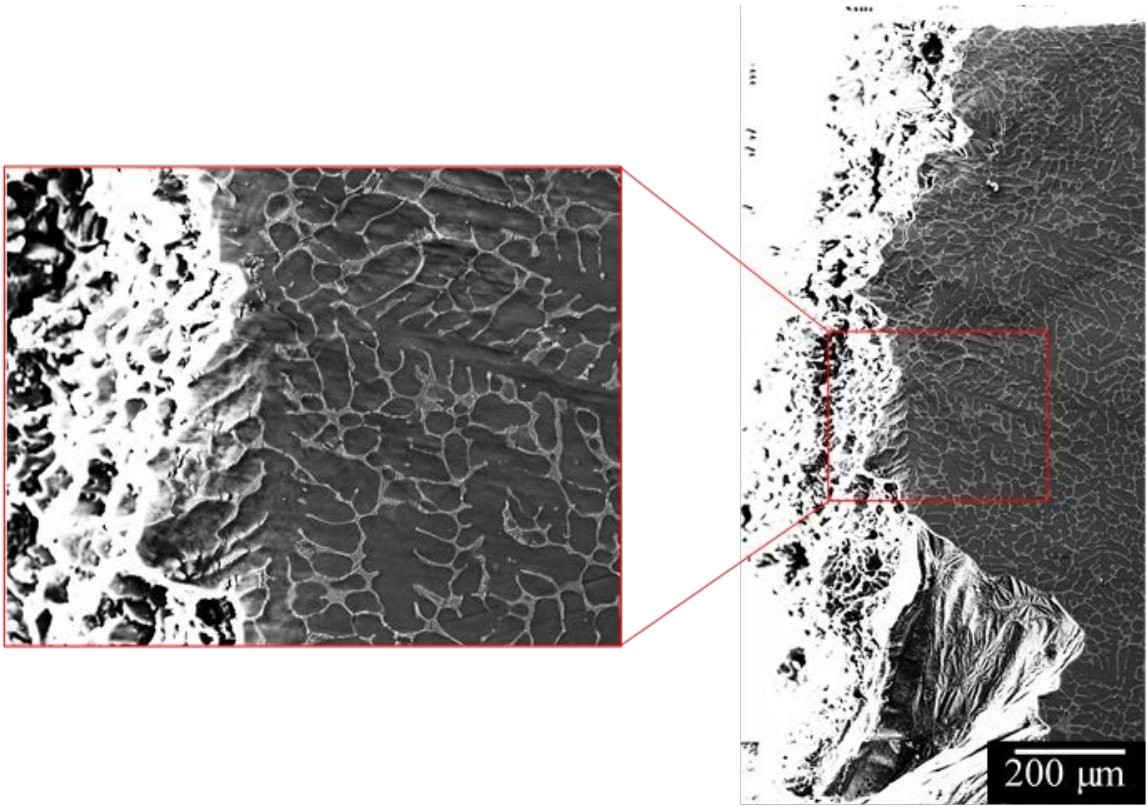

*Fig. 12. Sample fracture surface (deformed at 170 °C), crack growth is observed to follow the Laves phase network.*

This means that a high connectivity of the Laves phase network provides a preferential crack growth path, while the α-Mg grains in loosely connected networks might cause crack blunting causing increased deformability. This is in agreement to our earlier work [16] where we showed that an Mg-Al-Ca alloy with a loosely connected Laves phase network (C/A 0.3) has a higher tensile elongation of 5.1 % at 170 °C than an Mg-Al-Ca alloy with a highly connected Laves phase network (C/A 1.0) (tensile elongation of 3.6 % at 170 °C). However, the $\sigma_{0.2}$ and the UTS of the alloy with the highly connected Laves phase network ($\sigma_{0.2} \approx$ 72 MPa and UTS $\approx$ 113 MPa for alloy C/A 1.0) were higher than those of the alloy with a loosely connected Laves phase network ($\sigma_{0.2} \approx$ 58 and UTS $\approx$ 98 MPa for alloy C/A 0.3). Fig. 13 shows the Euler numbers of these two alloys revealing indeed the Euler number is lower for alloy C/A 1.0 than for alloy C/A 0.3. Similarly, it was shown in previous work that the Euler number of structures with high connectivity decreases and vice versa [36]. Kruglova et al. [36] showed for Al-Si alloys, that the Euler number can be effectively used to represent the connectivity of microstructural elements and reported that the strength of Al-Si alloys increased with decreasing Euler number, i.e. with higher connectivity of the intermetallic skeleton. Bugelnig et al. [59] showed that an increasing amount of Ni from 2 wt.% to 3 wt.% in an AlSi12Cu4Ni(2,3)Mg alloy increases the local connectivity, quantified by a more negative Euler number.

We therefore conclude that besides other microstructural parameters, such as crystal orientations, grain sizes, etc., the morphology and connectivity of the Laves phase network determines the mechanical properties of dual-phase Mg-Al-Ca alloys. Specifically, a highly connected Laves network is increasing the strength and also creep resistance on the one hand, but acts as preferential crack growth path reducing the formability of the alloy on the other hand. Therefore, for application of such alloys a compromise between (i) strength and creep resistance and (ii) ability to accommodate mechanical loads by plastic deformation has to be made. Further, as most of the cracks in the Laves phase are induced by strong slip localisation in the α-Mg phase, it is suggested that texture engineering to decrease the activation of basal slip might be used to design materials with a lower number of potential crack nucleation sites.

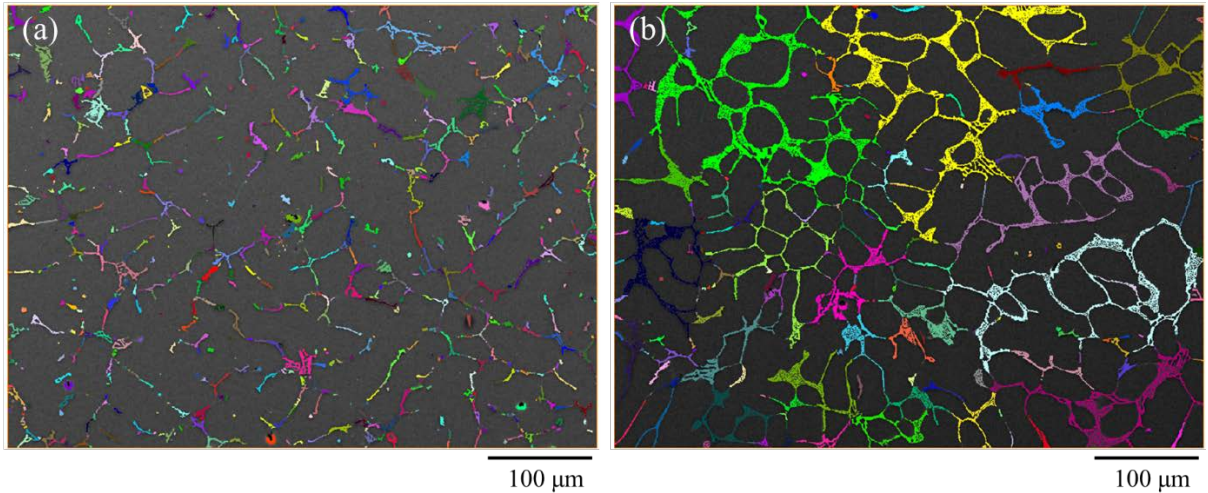

*Fig. 13. (a) Segmented Laves phases of an Mg-Al-Ca alloy with a Ca/Al ratio of 0.3, Euler Number: 681 and (b) an Mg-Al-Ca alloy with a Ca/Al ratio of 1.0, Euler Number: -2811*

**Conclusions**

We draw the following conclusions from this study:

1. A considerable amount of strain partitioning and localisation was observed at the microstructure level during elevated temperature tensile deformation using quasi in-situ tests with μ-DIC and:

    a. the strain is mainly concentrated in α-Mg grains with a high Schmid factor for basal slip or tensile twinning.

    b. strain localisations are further observed at the intersections of basal slip lines and deformation twins with α-Mg/Laves phase interfaces.

2. Quasi in-situ and bulk material analysis showed the initiation of micro damage in the Laves phase at the sites of maximum strain localisations, namely :

    a. intersections of slip lines in the α-Mg matrix with α-Mg/Laves phase interfaces.

    b. intersections of deformation twins with α-Mg/Laves phase interfaces and where deformation twins transfer across α-Mg/Laves phase interfaces.

3. Crack growth at large strains occurs preferentially along the Laves phase network.
4. The morphology and connectivity of the Laves phase network in Mg-Al-Ca alloys affect the mechanical properties: a high Laves phase connectivity causes high strength and creep resistance, but a low tensile ductility. On the other hand, microstructures with loosely connected Laves phase networks display low strength and creep resistance but higher tensile deformability.

**Acknowledgments**

M. Zubair gratefully acknowledges the financial support from the University of Engineering and Technology, Lahore Pakistan. Further, we are also very thankful to our colleagues, Mr. David Beckers, Mr. Thomas Burlet, Mr. Risheng Pei, Mr. Gerhard Schutz and Mr. Arndt Ziemons (alphabetic order) at the Institute for Physical Metallurgy and Materials Physics, RWTH Aachen for their help at various steps in this study.

4

55. Kusche, C., et al., *Large-area, high-resolution characterisation and classification of damage mechanisms in dual-phase steel using deep learning.* PLOS ONE, 2019. **14**(5): p. e0216493.
56. Yang, F., et al., *Crack initiation mechanism of extruded AZ31 magnesium alloy in the very high cycle fatigue regime.* Materials Science and Engineering: A, 2008. **491**(1-2): p. 131-136.
57. Bieler, T.R., et al., *Strain heterogeneity and damage nucleation at grain boundaries during monotonic deformation in commercial purity titanium.* JOM, 2009. **61**(12): p. 45-52.
58. Ng, B.C., et al., *The role of mechanical twinning on microcrack nucleation and crack propagation in a near-γ TiAl alloy.* Intermetallics, 2004. **12**(12): p. 1317-1323.
59. Bugelnig, K., et al., *Revealing the Effect of Local Connectivity of Rigid Phases during Deformation at High Temperature of Cast AlSi12Cu4Ni(2,3)Mg Alloys.* Materials (Basel), 2018. **11**(8).